\documentclass[referee,a4paper,12pt,traditabstract]{swsc}

\usepackage{graphicx}
\usepackage{txfonts}
\usepackage{subfigure}
\usepackage{epstopdf}
\usepackage{lineno}
\usepackage[authoryear,round]{natbib}
\usepackage[backref]{hyperref}
\usepackage{url}
\usepackage{multirow}
\usepackage{hyperref}

\hypersetup{colorlinks=true,citecolor=cyan,urlcolor=cyan,linkcolor=blue}

\begin{document}


   \title{Geomagnetic storm forecasting service StormFocus: 5 years online}

   \titlerunning{Geomagnetic storm forecast}

   \authorrunning{Podladchikova et al.}

\author{T. Podladchikova\inst{1}
          \and
          A. Petrukovich\inst{2}
					\and
					Y. Yermolaev\inst{2}}
					
   \institute{Skolkovo Institute of Science and Technology, Moscow, Russia\\
              \email{\href{mailto:t.podladchikova@skoltech.ru}{t.podladchikova@skoltech.ru}}
         \and
             Space Research Institute, Moscow, Russia\\
             \email{\href{mailto:a.petrukovich@cosmos.ru}{a.petrukovich@cosmos.ru}}
						}



  \abstract
   { Forecasting  geomagnetic storms is highly important for many space weather applications.
	In this study we review performance of the geomagnetic storm forecasting service StormFocus
	        during 2011--2016. The service was implemented in 2011 at
					\href{http://spaceweather.ru/content/extended-geomagnetic-storm-forecast}{SpaceWeather.Ru}
					and predicts the expected strength of geomagnetic storms as measured by $Dst$ index several hours ahead.
				  The forecast is based on L1 solar wind and IMF measurements and is updated every hour.
					The solar maximum of cycle 24 is weak, so most of the statistics are on rather moderate storms.
					We verify quality of selection criteria, as well as reliability of real-time input data in comparison
					with the final values, available in archives. In real-time operation
					87\% of storms were correctly predicted while the reanalysis running on final OMNI data predicts
					successfully 97\% of storms. Thus the main reasons for prediction errors are discrepancies between real-time
					and final data ($Dst$, solar wind and IMF) due to processing errors, specifics of datasets.}


   \maketitle

\section{Introduction}
When the southward interplanetary magnetic field (IMF) embedded in the solar wind affects the Earth’s
magnetic field during several hours, a substantial energy is transferred into the magnetosphere.
The entire magnetosphere becomes disturbed and this
state is characterized as "geomagnetic storm" \citep[e.g.,][]{GonzalezWhatIsGeomagneticStorm1994}.
The common measure of intensity of a geomagnetic storm is $Dst$ geomagnetic index --- a depression of
equatorial geomagnetic field, associated primarily with the magnetospheric ring current.
The strength of geomagnetic storm
is characterized by the maximum negative value of the index, which is denoted hereafter as $Dst_p$.

Operational forecasting of geomagnetic storms is highly important for space weather applications.
The geomagnetic storms can be predicted qualitatively, when the large scale events on the Sun (flares, CMEs or coronal holes) are detected. The propagation time of solar wind from
Sun to Earth is about 1--5 days, creating the natural time interval for such predictions.
However, the magnetic structure of an interplanetary perturbation (magnetic cloud or stream interface),
in particular IMF $B_z$ profile (hereafter GSM frame of reference is used), which is of prime importance for the magnetospheric dynamics, can not be determined for now from solar observations.
Thus, detailed forecast of a storm strength several days in advance is currently impossible.

Quantitative geomagnetic predictions can be performed using the measurements of interplanetary magnetic field
and solar wind by a spacecraft in the libration point L1. The difference in the propagation
time from L1 to Earth between radio signal and solar wind is typically one hour, providing the
chance for the short-term forecast. Intrinsic solar wind and IMF variations
can deteriorate the quality of such a forecast, changing the input on the way from L1 to the magnetosphere.
However, as shown in \citet{Petrukovich2001}, the relatively large-scale storm-grade interplanetary disturbances
are well preserved between L1 and Earth, while the smaller-scale substorm-grade disturbances indeed can change substantially.
The gap between two currently available methods (solar and L1) might be filled, in particular, by developing
the methods of advance forecasting  several hours ahead of available solar wind data.

The modeling of $Dst$ index is relatively well performed using IMF and solar wind as input.
Current approaches can be, in general, divided into several groups.
The first variant is based on the statistical dependence of the $Dst$ minima during storms on
geoefficient solar wind parameters (southward magnetic field or coupling functions)
\citep[e.g.,][]{{Akasofu1981},{Petrukovich2000}, {Yermolaev2002},
{GonzalezEcher2005}, {Yermolaev2005}, {KaneEcher2007}, {Mansilla2008},{EcherGonzalez2008},{Yermolaev2010},
{Echer2013},{Rathore2014}}.

The second approach is based on a first-order differential equation of $Dst$ index evolution
depending on geoeffective solar wind parameters \citep{Burton1975}. This approach with the
later modifications \citep{{OBrien2000b}, {OBrien2000a}, {Siscoe2005},{Nikolaeva2014}}
proved to be very successful and helps to model dynamics of $Dst$
index in detail.

The  third group uses various black box-type statistical models,
relating the solar wind and $Dst$ index: artificial neural networks, nonlinear auto-regression schemes, etc.
\citep{{Valdivia1996}, {Vassiliadis1999}, {Lundstedt2002}, {Temerin2002},
{Temerin2006}, {Wei2004}, {Pallocchia2006}, {Sharifie2006}, {Zhu2006},
{Zhu2007}, {Amata2008}, {Boynton2011}, {Boynton2013}, {Revallo2014}, {Caswell2014}, {Andriyas2015}}. These models with rather complex structure are capable to extract information about the process without any prior assumptions.
A detailed comparison of several such models was done for large storms ($Dst <$--100 nT) by \citet{Ji2012}.

Finally some physics-based magnetospheric models are capable in modeling $Dst$ \citep[e.g.,][]{Katus2015}.
\citet{Rastatter2013} compared statistical and physics-based models and also presented a nice review of approaches.

Operational use of such $Dst$ models with the L1 real-time solar wind
naturally provides a forecast with the lead time of the order of one hour.
Forecasting $Dst$ several hours ahead (of available solar wind) is a more challenging topic.
One can generate such a forecast as a simple extension of the ``black-box'' approach
described above. \citet{WuLundstedt1997} and \citet{Sharifie2006} used
the neural network and local linear neuro-fuzzy models, however these results were
not presented in sufficient details. A similar approach was used to predict $Kp$ ahead of solar wind by \citet{Wing2005}. \citet{Bala2012} used an artificial neural network to forecast several indices via the empirical estimate of the Earth's polar cap potential. Frequently authors also discuss the "one step ahead" forecast \citep{Revallo2014}.

Considering approaches to multi-hour forecasts it is helpful
to keep in mind the following aspects:
(1) Multi-hour forecasts require some supposition on the expected solar wind behavior.
In the ``black-box'' type models this information is hidden from the user. However  it might
be essential to have full control on assumption about solar wind input.
(2) Since evidently some certainty in the input is lost, it is
natural to step back also in the forecast details.
For example, one can formulate results in terms of thresholds.
(3) The continuous $Dst$ timeline is dominated by geomagnetically quiet intervals,
thus it might be more reasonable to check the model quality only on storm events.

Several approaches were suggested, which essentially follow these ideas.
\citet{Mansilla2008} showed the straight relation  between the peak $Dst_{p}$ and the
peak value of the solar wind velocity $V$, and determined the time delay between $Dst_{p}$
and maximum negative $B_{z}$ to be about several hours.
However, this particular method needs statistical justification of specific forecasts.
\citet{Chen2012} used bayesian model to forecast thresholds of storm strength,
using the past statistics of amplitude and duration of southward IMF $B_z$, in particular for magnetic clouds.

Often the past measured  $Dst$ indices are also used as input,
as currently the quick-look $Dst$ is available almost immediately. In a case of delay
in quick-look $Dst$, the index can be reasonably well reconstructed
using solar wind \citep[e.g.,][]{{Lundstedt2002}, {Wei2004}, {Sharifie2006}, {Zhu2007}}.

In the first publication \citep{PodladchikovaPetrukovich2012} we developed the prediction technique of the geomagnetic storm peak strength $Dst_{p}$
at the first relevant signs of storm-grade input in the solar wind.
It uses the extrapolation of the Burton-type $Dst$ model with the constant solar wind input to provide
the forecast several hours ahead of available solar wind data.
The method essentially relies on the relative persistency of
large-scale storm-grade solar wind and IMF structures and on the cumulative
nature of $Dst$ index, partially integrating out input variations. In fact, the stronger
is the expected storm, the easier is such forecast.
On the basis of the proposed technique a new online geomagnetic storm forecasting service  was implemented
in 2011 at \href{http://spaceweather.ru/content/extended-geomagnetic-storm-forecast}{SpaceWeather.Ru}. Since
2017 we adopted the name StormFocus.
In this study we review performance of StormFocus during more than five years of operation 2011--2016 and
verify the algorithm thresholds and other calculation details. The solar maximum of the cycle 24 is rather weak,
so most of new statistics are rather moderate storms.
We also analyze the origins of the observed errors. Besides some imperfectness of the algorithm, another
major real-time error type proved to be related with the missing data and the calibration-related differences between
the real-time and the final data. These latter errors are often not recognized, but actually
account for a significant part of the forecast uncertainty.

More specifically, all models are designed on the final quality
solar wind data, usually from the OMNI dataset, which appear with the delay of several months.
OMNI data are taken from one of the several available spacecraft, verified,
and shifted to Earth with a relatively complicated algorithm.
The real-time data provided by NOAA are unverified, not shifted and come only from ACE (currently also from DSCOVR).
As concerns $Dst$, the
final index is available until 2010 (as of beginning 2017), while the
provisional one is for 2011--2016.
It might be quite different from  the real-time $Dst$, which is used in the model
(as the previous value) or which  is compared with the forecast result in real-time.
Our archived forecast history includes also input data and  allows us to analyze these errors in detail,
comparing real-time algorithm performance with the reanalysis on final OMNI and ACE/Wind spacecraft data, as well
as directly comparing real-time and final input data (see also end of Sec. 2).

In the section~2 we explain our method. Section~3 describes the statistics of the forecast quality over the period of service operation
July 2011 -- December 2016 (hereafter --- the test period), including both real-time results and reanalysis using final data.
Section 4 analyzes the actual error sources, including the differences between real-time
and final data. Section 5 concludes with Discussion.

\section{Methodology}
The geomagnetic storm forecasting service StormFocus was implemented in 2011 at the website of Space Research Institute, Russian Academy of Sciences (IKI, Moscow) \href{http://spaceweather.ru/content/extended-geomagnetic-storm-forecast}{SpaceWeather.Ru}.
The full details of the prediction algorithm were in our first publication \citep{PodladchikovaPetrukovich2012} and are now included as Appendix~A.
ACE real-time data, obtained from NOAA SWPC and real-time  quicklook $Dst$ are used as input.
The prediction algorithms were initially tuned on historical data from 1995 to 2010.
During this period 97 storms with ($Dst_{p}<-100$~nT) and 317 storms  ($-100<Dst_{p}<-50$~nT)  were registered.
The DSCOVR real-time data have been used as input, starting in October 2017.
Data are averaged (boxcar) at round hours and forecasts are computed every hour. 
The forecasts are archived and available at
\href{http://www.iki.rssi.ru/forecast/data/Archive}{http://www.iki.rssi.ru/forecast/data/Archive}.

The best predicted are relatively large storms with sharp turns to strong southward IMF $B_z$ (essentially
with large $VBs$), identified with some empirical thresholds explained in Appendix~A. For such storms StormFocus provides the
upper and lower limits of the future peak $Dst_{p}$. These storms are called ``sudden'' in the
terminology of \citet{PodladchikovaPetrukovich2012}. All other storms are called ``gradual'' storms
since  $VBs$ increases with no clear step. For them the service provides the single prediction of
$Dst_{p}$. Our forecast is routinely updated every hour and the maximal (the most negative) prediction is kept active until storm end is signaled.

Such forecasts of storm peak values proved to be rather reliable.
Errors were at the moderate level of $\sim$10\%.
Predictions were issued on average
5--6 hours before the actual peak was registered.
Accuracy of predicting the time of $Dst_{p}$ turned out to be much worser than that of $Dst_{p}$ itself, 
thus time of storm maximum  is not forecasted in our tool. Particular time of $Dst_{p}$, especially for “gradual” storms, is likely influenced by some transient variations in solar wind or geomagnetic activity, which is 
in contradiction with our main hypothesis on input persistence.

\begin{figure}
\centerline{\includegraphics[width=1\textwidth,clip=]{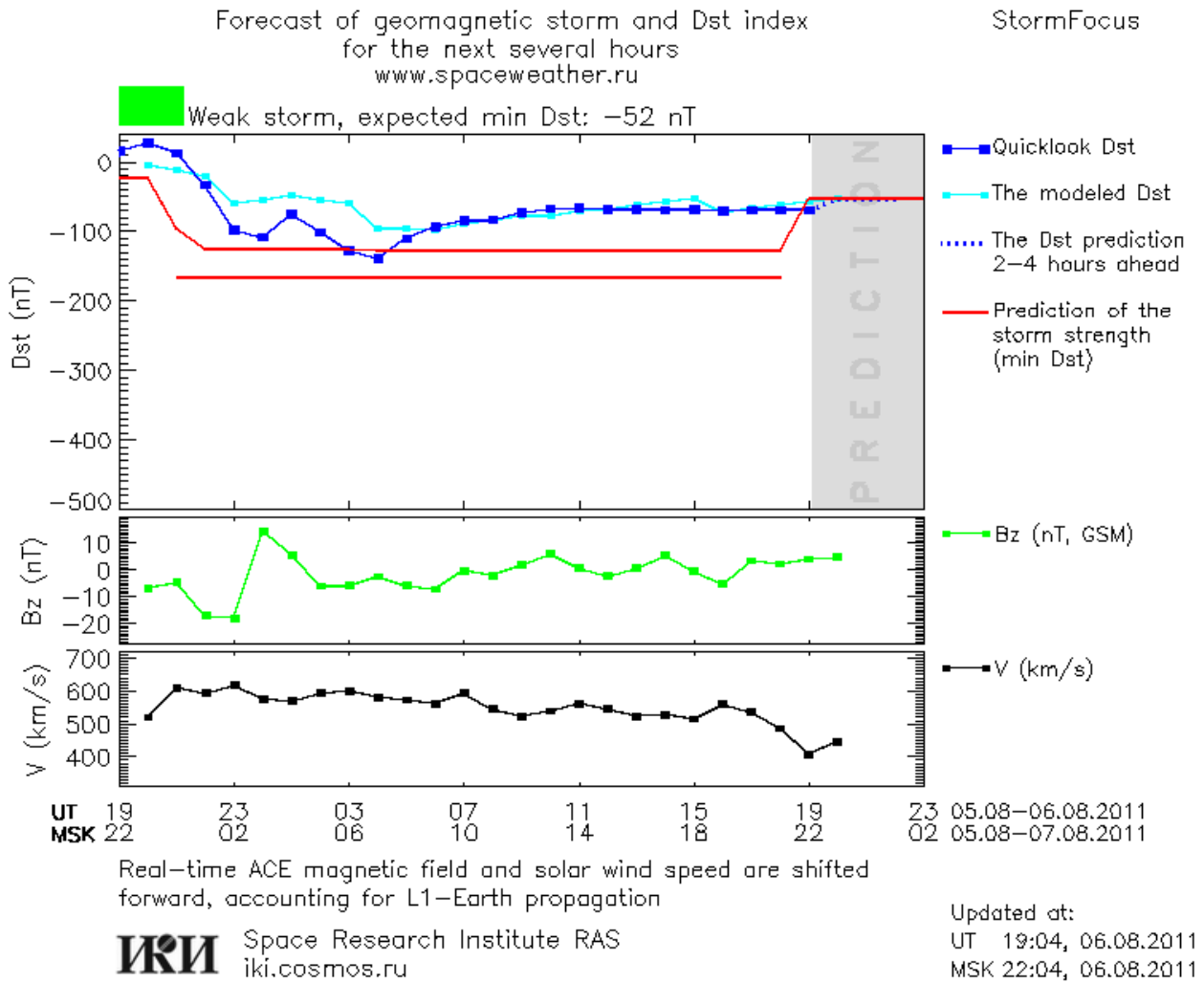}}
\caption{Layout of the StormFocus forecast web page for the geomagnetic storm on 6 August 2011.}
\label{fig1}
\end{figure}

Figure~\ref{fig1} shows the StormFocus forecast page layout for the geomagnetic storm on 6 August 2011 with the peak
$Dst_{p}=-138$~nT. It includes the panels with quick-look  $Dst$ (blue), modeled $Dst$ (cyan) using Equations~(\ref{Eq_Burton}--\ref{Eq_Dst_cor}), and predictions (red),
with IMF $B_z$ (green), and solar wind velocity $V$ (black), as well as the verbal forecast.
Here the real-time ACE $B_{z}$ and the solar wind $V$ are ballistically shifted forward,
accounting for the L1-Earth propagation. At 21:00~UT, 5 August 2011 the ``sudden'' storm was predicted with the limits
from --166 to --94~nT. This prediction of peak magnitude was provided 7 hours
before it was actually registered, at a moment when  $Dst$  was still positive (14 nT). Note, that the verbal warning
of --52 nT storm, shown on the top of picture, corresponds to the later time 9:05~UT, when this screen-shot was generated.

In this investigation we estimate the performance of StormFocus operation during 2011--2016.
Similar to our first publication we aim at a reliable quantitative forecast.
A standard approach to forecast
errors specifies
false warnings (errors of the first kind), misses (errors of the second kind), as well as true negatives (correct prediction of quiet interval).
Since we issue forecast of storm peak magnitude every hour
with the floating waiting time to the real peak, there is no consistent definition
of the true negative. Moreover, a prediction of a quiet interval
several hours ahead has no solid physical basis in solar wind data, since there
is no guarantee, that some strong disturbance will not appear
``next hour'', especially when Sun is active.
Thus predictions of geomagnetically
quiet intervals (though it may be of interest to some applications) are not
a part of our forecast. We aim to
predict only storms with  $Dst<-50$~nT.

In our scheme true misses are very unlikely, since L1 spacecraft
is a reliable upstream monitor and no storms occur without
corresponding interplanetary disturbance. However,
in some cases the first satisfactory forecast could be issued rather
late, well inside the developing storm. Our prime goal was to
minimize false warnings (predictions of too large storms).

During quality checks
we first identify all storms (below —50 nT) and their $Dst_p$. 
Then we search for the 
earliest correct (within 25\%) forecast before each observed peak.
If all forecasts (before actual $Dst_p$) are more than 25\% weaker 
(less negative) than $Dst_{p}$, the storm is considered as missed
(though the forecast still may reach ``correct'' amplitude later than the actual
$Dst_{p}$ was observed). If there is no correct forecast before 
the actual peak, but there is prediction 25\% stronger (more negative) than the peak $Dst_{p}$, the storm amplitude is considered overestimated (false warning). To check the usefulness of the forecast we also determine the advance time, when the successful forecast was issued (relative to actual $Dst_{p}$), and $Dst$ values at the moment of forecasts.

One more characteristic is the maximal prediction issued within a storm.
The third variant of an error is identified, when
this maximal prediction happens  
after the first correct forecast and overestimates true $Dst_p$ 
by more than 25\%. This is called in the following ``overestimated maximum prediction''.

We verify the general prediction accuracy in three variants with respect to the used data.
The first variant is ``true comparison'' of the forecast
(using ACE real-time solar wind, IMF and previous quicklook $Dst$) with the
storm magnitude as measured by quicklook  $Dst_{p}$.
However, from a point of view of a general user, it might be more correct to compare the
forecast with  actual (final) $Dst$. This is the second variant of our comparison.
For the third variant we rerun the algorithm using final solar wind, IMF and $Dst$ from OMNI
 and compare with the storm magnitude as measured with final (or provisional) $Dst_{p}$.
This latter approach is also known as reanalysis.

For the ACE real-time solar wind, IMF and quicklook $Dst$ data we use our own archive for 2011--2016,
which was filled in the course of operation.
Final solar wind and IMF data are taken from OMNI dataset (shifted and merged data) as well as from CDAWeb for individual ACE and Wind spacecraft (not shifted original data). The final $Dst$ is available only before 2011, thus for the later time we use provisional $Dst$  (see also discussion in Sec.4).
In the latter text (if not explicitly stated oppositely)
we use the term ``final'' to characterize OMNI solar wind and provisional or final $Dst$ as opposed to
the real-time data.

\section{Forecast statistics}
In this section we present the statistics of the forecast quality over the period of StormFocus operation July 2011 -- December 2016, as well as the reanalysis using the final data. The $Dst_{p}$ of relatively large ``sudden'' storms with the sufficiently sharp increase of $VBs$ (see criteria in Appendix~A (\ref{Eq_Jump1})~and~(\ref{Eq_Jump2}) is forecasted with the lower and upper magnitude limits.
During the test period the sudden storm criteria in real-time were activated only for 10 storms.
However with the final data reanalysis  there were 23 ``sudden'' storms.
These storms were weaker than that for 1995--2010 (53 storms)
used by \citet{PodladchikovaPetrukovich2012} to design the prediction algorithms.
Over the last 5 years only three  storms were below  --150~nT, while for 1995--2000
50\% of such storms were with peak $Dst_{p}<-150$~nT. The ``sudden'' storm forecast works
best for larger storms, so the conditions during last five years were not favorable in this sense.

\begin{figure}
	\centerline{\includegraphics[width=1\textwidth,clip=]{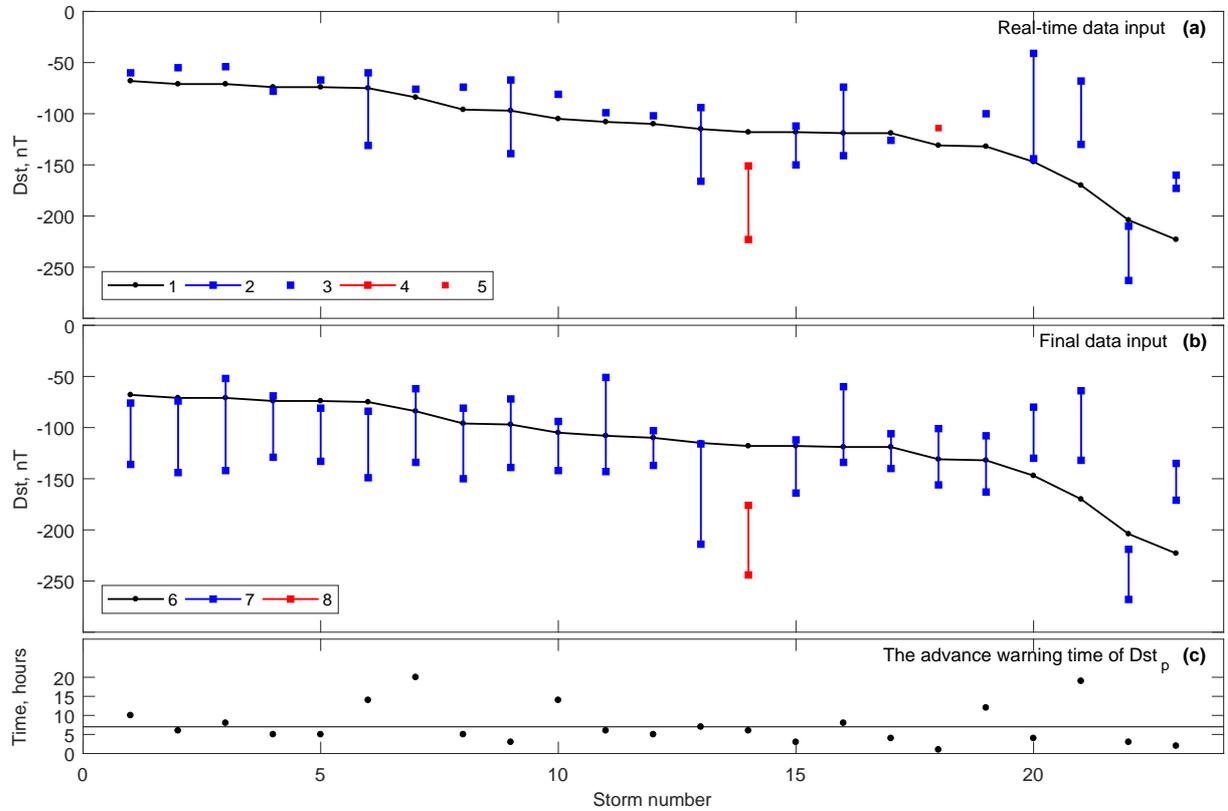}}
	\caption{Forecast statistics  for ``sudden'' geomagnetic storms.
			(a) Storm predictions using real-time data input.
			(1) final $Dst_{p}$;
			(2) the earliest prediction of both upper and lower limits of $Dst_{p}$ within 25\% of actual storm magnitude;
			(3) the earliest prediction of $Dst_{p}$ with 3-hour forecast (not ``sudden'' in real-time);
			(4) and (5) the predictions of $Dst_{p}$, which were out of 25\% range from actual $Dst_{p}$.
			(b) Storm predictions using final data input.
			(6) final $Dst_{p}$;
			(7) the earliest prediction of both upper and lower limits of $Dst_{p}$ within 25\% of actual storm magnitude;
		    (8) the predictions of $Dst_{p}$, which were out of 25\% range from actual storm magnitude.
			(c) The advance warning time (in hours) of the $Dst_{p}$ forecast using real-time input.}
	\label{fig2}
\end{figure}

Figure~\ref{fig2}a shows the forecast statistics for 23 ``sudden'' geomagnetic storms (as determined in final data). We use real-time data input
compared with the final geomagnetic index, and  storms are ordered by the final $Dst_{p}$ (black line).		
The  bars in Figure~\ref{fig2}a show the forecast for 10 events, for which the ``sudden'' criterion was ``on'' in real-time, and both upper and lower forecast limits were
calculated. The remaining 13 storms (marked by single points) were characterized as ``gradual'' in real-time, for which only one-point forecast (Appendix~A) was produced. The red bar  and red point indicate unsuccessful forecasts for two events. Thus, StormFocus service issued a successful forecast using the real-time input data for 21 storms, though in 9 cases it was provided with two limits and in 12 cases - with a single point. Storm forecasts were produced when the $Dst$ index was on average weaker by 75\% than the final $Dst_{p}$, and in some cases the warnings were issued, when $Dst$ index was positive (not shown here). The advance time of the real-time forecasts  was 1--20 hours with the average value of about 7 hours (Figure~\ref{fig2}c).

Figure~\ref{fig2}b shows the results of reanalysis on final data for 23 geomagnetic storms. The blue bars show the successful forecast of upper and lower limits of $Dst_{p}$ (black solid line) based on final OMNI data for 22 storms. The only red bar gives unsuccessful ``overestimate'' forecast, which was more than 25\% stronger (more negative) than actual $Dst_{p}$. Note, that this forecast was still produced in advance, and the storm was in fact detected, though the strength was overestimated. To check the usefulness of forecast we also determined the final $Dst$ index at the issue time of $Dst_{p}$ forecast. On average it is weaker by 84\% than the final peak $Dst_{p}$.

Storms with no ``sudden'' criterion are categorized as gradual and
their magnitude $Dst_{p}$ is predicted with the single number based on the three-hour forecast $\overline{Dst}$ (Appendix~A).
Five ``gradual'' storms had $Dst_{p} <$--100 nT and 92 storms  had $-100<Dst_{p}<-50$~nT.
The classification of storms is made using more reliable final $Dst$.
Additionally, 33 storms had real-time $Dst_{p}<-50$~nT, but final   $Dst_{p}>-50$~nT.
These storms were also included to the forecast statistics.

\begin{figure}
	\centerline{\includegraphics[width=1\textwidth,clip=]{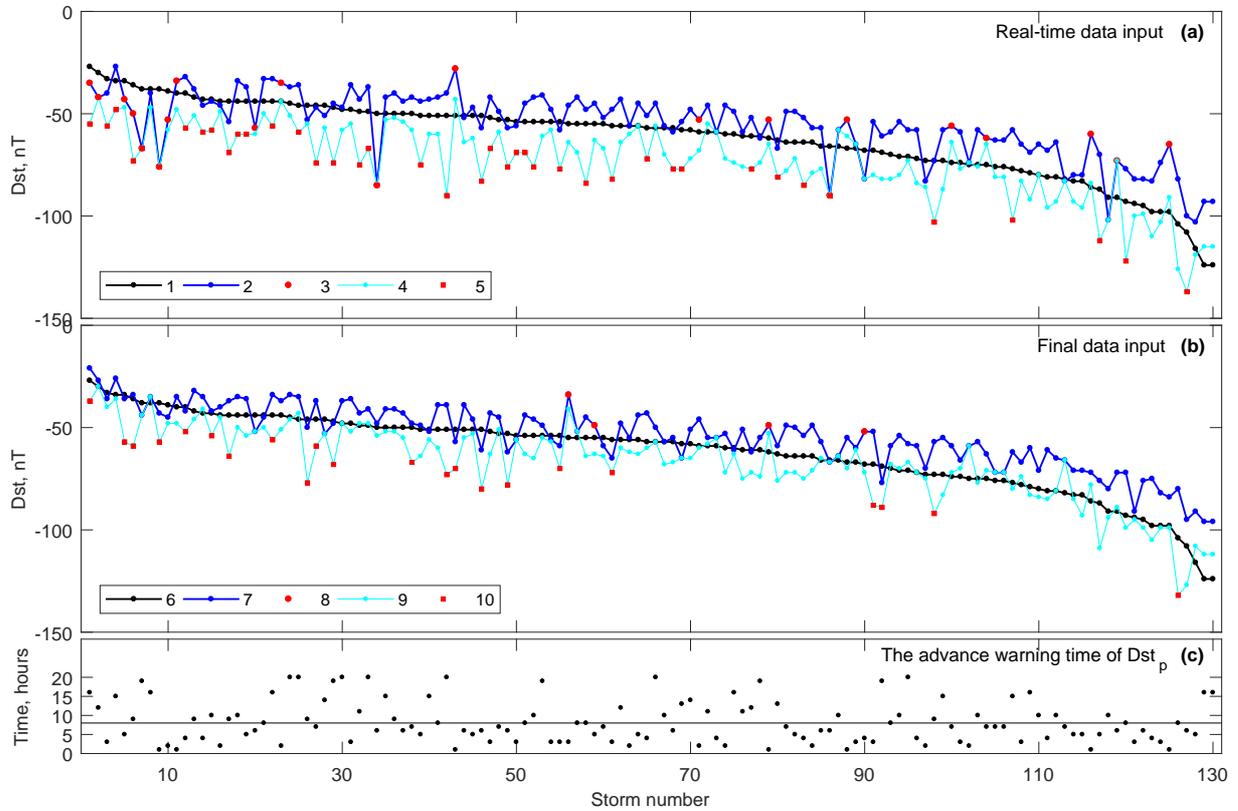}}
	\caption{Forecast statistics for ``gradual'' geomagnetic storms.
			(a) Storm predictions using real-time data input.
			(1) final $Dst_{p}$;
			(2) the earliest prediction of $Dst_{p}$ within 25\% of actual storm magnitude;
			(3) the predictions of $Dst_{p}$, which were out of 25\% range;
			(4) maximal predictions of $Dst_{p}$ obtained during the storm development;
			(5) maximal predictions out of 25\% range.
			(b) Storm predictions using final data input.
			(6) final $Dst_{p}$;
			(7) the earliest prediction of $Dst_{p}$ within 25\%;
			(8) the predictions of $Dst_{p}$, which were out of 25\% range;
			(9) maximal predictions of $Dst_{p}$ obtained during the storm development;
			(10) maximal predictions out of 25\% range.
			(c) The advance warning time (in hours) of the $Dst_{p}$ forecast using real-time input.}
	\label{fig3}
\end{figure}

Figure~\ref{fig3} presents the statistics of gradual storms relative to final $Dst_{p}$.
With the real-time forecast, the earliest ``successful'' prediction (blue line in Figure~\ref{fig3}a) of final $Dst_p$ (black line) was produced for 109 out of 130 cases. The outliers are marked with red points, including 10 ``overestimated'' events, and 11 missed predictions. Cyan line shows the maximal predictions issued later within a storm. The number of overestimated maximal predictions was 37 (red points). Note that for these
 events the earliest forecast was successful.
Mainly small storms were overestimated, when errors of a baseline are relatively more important. The storm forecasts were produced when the $Dst$ index was on average weaker by 57\% than the final $Dst_{p}$, confirming the usefulness of prediction service. The advance time of the real-time forecasts was 1--20 hours with the average value of about 8 hours (Figure~\ref{fig3}c).

With the reanalysis using final OMNI data (Figure~\ref{fig3}b), 126 out of 130 storms were successfully predicted. The blue line gives the earliest prediction, which is in 25\% range from final $Dst_{p}$. The red points show unsuccessful missed forecast at four events.
Cyan line shows the maximal predictions of $Dst_{p}$ issued later within a storm. The number of overestimated maximal predictions with the final data is about a half of that with real-time input (in comparison with Fig. 3a) --- just 21 storms. $Dst$ at the issue time of forecast was on average weaker by 54\% than the final OMNI $Dst_{p}$.

Table~\ref{table1} summarizes the forecast statistics.
\renewcommand{\tabcolsep}{3pt}
\begin{table}
	\caption{
		Forecast statistics for: ``RR'' - real-time quicklook $Dst_{p}$ prediction using real-time input;
		``FF'' - final $Dst_{p}$ prediction using final input;
		``RF'' - final $Dst_{p}$ prediction using real-time input.
		}
	\label{table1}
	\begin{tabular}{cccccc}     
		\hline \\[-1.0em]        
		& \parbox{1.5cm}{\centering Number \\of storms}
		& \parbox{1.5cm}{\centering Successful \\forecast}
		& \parbox{2.8cm}{\centering Overestimated}
		& \parbox{2cm}{\centering Missed}
		& \parbox{2.8cm}{\centering Overestimated \\maximal\\ prediction} \\[0.5cm]
		\hline \\ [-2ex]
		&  & RR\,\,FF\,\,RF & RR\,\,FF\,\,RF  & RR\,\,FF\,\,RF & RR\,\,FF\,\,RF \\ 		
		\parbox{4.1cm}{\centering Sudden storms \\Sharp increase of $VBs$}         & 23  & 21 \enspace 22 \enspace 21   & 1\enspace\enspace 1 \enspace\enspace 1   &1 \enspace\enspace -\enspace\enspace 1           &2\enspace\enspace 3 \enspace\enspace 5    \\[0.5cm]
		\parbox{4cm}{\centering Gradual  storms \\$Dst_{p} \leq -100$ nT}   & 5   & 3 \enspace\enspace 5 \enspace\enspace 5  & 1 \enspace\enspace - \enspace\enspace -  & 1 \enspace\enspace - \enspace\enspace -            & - \enspace\enspace 1 \enspace\enspace 1   \\[0.5cm]
		\parbox{4.1cm}{\centering  Gradual storms \\$-100<Dst_{p}<-50$ nT}          & 92  & 78 \enspace 88 \enspace 81  & 2 \enspace\enspace - \enspace\enspace 2   & 12 \enspace 4 \enspace\enspace 9          & 7 \enspace 10 \enspace 22    \\[0.5cm]
		\parbox{4.2cm}{\centering  Gradual storms \\Final $Dst_{p}>-50$ nT, \\Quicklook $Dst_{p}<-50$ nT }                                           & 33  & 27 \enspace 33 \enspace 23 & 1 \enspace\enspace - \enspace\enspace 6   & 5 \enspace\ - \enspace\enspace 4           & 2 \enspace 10 \enspace 14    \\[0.5cm]
		\hline \\
		Total                                                                     & 153 & 129 \enspace 148 \enspace 130 & 5 \enspace\enspace 1 \enspace\enspace 9  & 19 \enspace 4 \enspace\enspace 14  & 11 \enspace 24 \enspace 42  \\
		\hline
	\end{tabular}
\end{table}
It includes the three variants of the forecast run: real-time $Dst_{p}$ prediction using real-time input (``RR''),
 final $Dst_{p}$ prediction using final input (``FF''),
 final $Dst_{p}$ prediction using real-time input (``RF'').
``Successful forecast'' gives the number of storms with the earliest forecast of peak $Dst_{p}$ within 25\% of the actual
storm magnitude.
``Overestimate'' and ``missed'' represent the number of unsuccessful forecasts. ``Overestimated maximal prediction'' shows the number of overestimating forecasts issued later within a storm (when the earliest forecast was successful).

As it is clear from Table~\ref{table1}, over the test period
our prediction algorithm performed best for the reanalysis (final input compared with final $Dst_{p}$), providing 148 successful predictions of $Dst_{p}$ out of 153 storms.
Real-time forecast compared with final $Dst_{p}$ resulted in 130 successes.
Real-time forecast compared with real-time quicklook $Dst_{p}$ had 129 successes.
Thus the usage of real-time input IMF, solar wind, and $Dst$ decreased the quality of the earliest forecasts
for 21 storms as compared with the final input.

In addition there were events with overestimation of maximal predictions
issued later within a storm (when the earliest forecast is successful). The worst quality
with respect to this criterion has the real-time forecast as compared with final index (RF variant).
The number of errors was almost twice larger than for RR variant, and almost four times larger than for FF variant (42 compared with 24 and 11).
The most of errors were for smaller storms (left side of Fig. 3). Most likely the main reason is the baseline difference  between real-time and final $Dst$. In the RF-variant the forecast is computed with the real-time quicklook Dst index, while comparison is with final $Dst$.

All possible outcomes for the forecasts of $Dst_{p}$ can be described by the number of hits (successful forecast), misses, and false alarms. We can calculate the following verification measures such as the probability of detection (POD) and the ratio of overestimated forecasts (ROF) (Table 2).
\begin{equation} \label{Eq_POD}
POD=\frac{hits}{hits+misses}
\end{equation}
\begin{equation} \label{Eq_ROF}
ROF=\frac{overestimated \quad forecasts}{hits+overestimated \quad forecasts}
\end{equation}

\begin{table}
	\caption{
	The probability of detection (POD) and the ratio of overestimated forecasts (ROF) of the earliest and maximal forecasts for the three types of the forecast run:
	``RR'' - real-time quicklook $Dst_{p}$ prediction using real-time input;
	``FF'' - final $Dst_{p}$ prediction using final input;
	``RF'' - final $Dst_{p}$ prediction using real-time input.}
	\label{table_POD_ROF}
	\begin{tabular}{cccc}     
		\hline \\[-2ex]        
		    & RR 					     & FF 						 & RF \\
		\hline \\[-2ex]
		\parbox{1.5cm}{\centering POD} & $\frac{129}{129+19}=0.87$  & $\frac{148}{148+4}=0.97$  &  $\frac{130}{130+14}=0.9$  \\[0.7cm]
		\parbox{1.5cm}{\centering ROF \\earliest forecast} & $\frac{5}{129+5}=0.04$     & $\frac{1}{148+1}=0.007$   &  $\frac{9}{130+9}=0.06$  \\[0.7cm]
		\parbox{1.5cm}{\centering ROF \\maximal forecast} & $\frac{11}{129+11}=0.08$     & $\frac{24}{148+24}=0.14$   &  $\frac{42}{130+42}=0.24$  \\[0.7cm]
		\hline
	\end{tabular}
\end{table}

The highest probability of detection (0.97) and the lowest ratio of overestimated forecast (0.007) is for FF reanalysis variant. The probability of detection decreased to 0.87 and the ratio of overestimated forecast increased to 0.04 for the real-time quicklook $Dst_{p}$ prediction using real-time input (RR). The numbers for RF variant remain similar to that for RR. Overestimates of the maximal forecast after successful warning are at the level of 24\% for RF variant and about 10\% for RR and RF variants.

Among errors in Table~\ref{table1}, there were in total 49 storms with some forecast errors in real-time and correct forecast with final data. These 49 errors can be definitely ascribed to real-time data errors. In the next section we analyze in detail the reasons of these differences.

\section{Analysis of reasons for the decrease of the forecast quality using real-time input}
Forecasting services use data available in real-time, which
can be substantially different from the final calibrated data appearing later
in the scientific archives. However, the forecast algorithms of $Dst$ are designed using the final data.
The forecast errors due to this factor are often overlooked, but can be
equally important as those due to an algorithm imperfection.
Our analysis in Section~3 shows that the number of the forecast errors is indeed much
smaller if the final quality data are used. In this section we analyze
specific sources of our forecast errors and the differences between real-time and final data in general.

\begin{figure}
	\centerline{\includegraphics[width=1\textwidth,clip=]{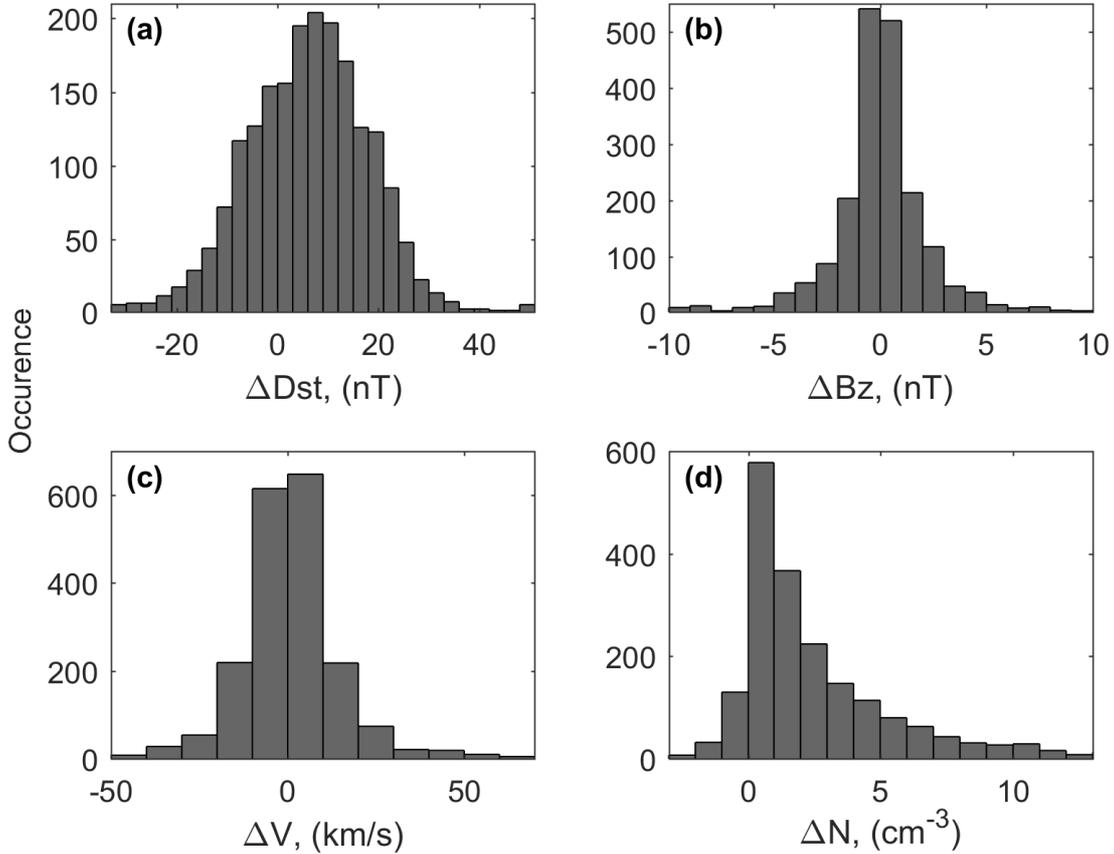}}
	\caption{The deviation of hourly averaged final data (OMNI) from hourly averaged real-time data for disturbed
		periods $Dst<-50$ nT over the period of July 2011 to December 2016. (a) $Dst$ index (nT).
		(b) Southward IMF $B_{z}$ (nT). (c) Solar wind velocity $V$ (km/s). (d) Solar wind density $N$ (cm$^{-3}$).}
	\label{fig4}
\end{figure}

Figure~\ref{fig4} shows the difference between the hourly averaged final data and real-time ACE and Dst values for the disturbed periods $Dst<-50$~nT over July 2011 -- December 2016.
 $Dst$ index is mostly overestimated, as quick-look index available in real-time is smaller (i.e., more negative) than the provisional $Dst$ (67\% of points) and for more than half of cases the difference exceeds 10~nT (Figure~\ref{fig4}a). This effect is responsible for the half of events under analysis (28 out of 49, Table~\ref{table2}, detailed description is below).

It should be noted additionally that quicklook $Dst$ is usually updated
approximately half an hour after the first appearance. We do not
have in hand the statistics of this change. Several years later provisional $Dst$ is
replaced by final $Dst$, and the scatter between these two index types
is a factor of 2-3 smaller, than that for the real-time --- provisional pair.
Also, the final $Dst$ is on average again weaker (more positive) than the provisional index by few nT.

\begin{figure}
\centerline{\includegraphics[width=1\textwidth,clip=]{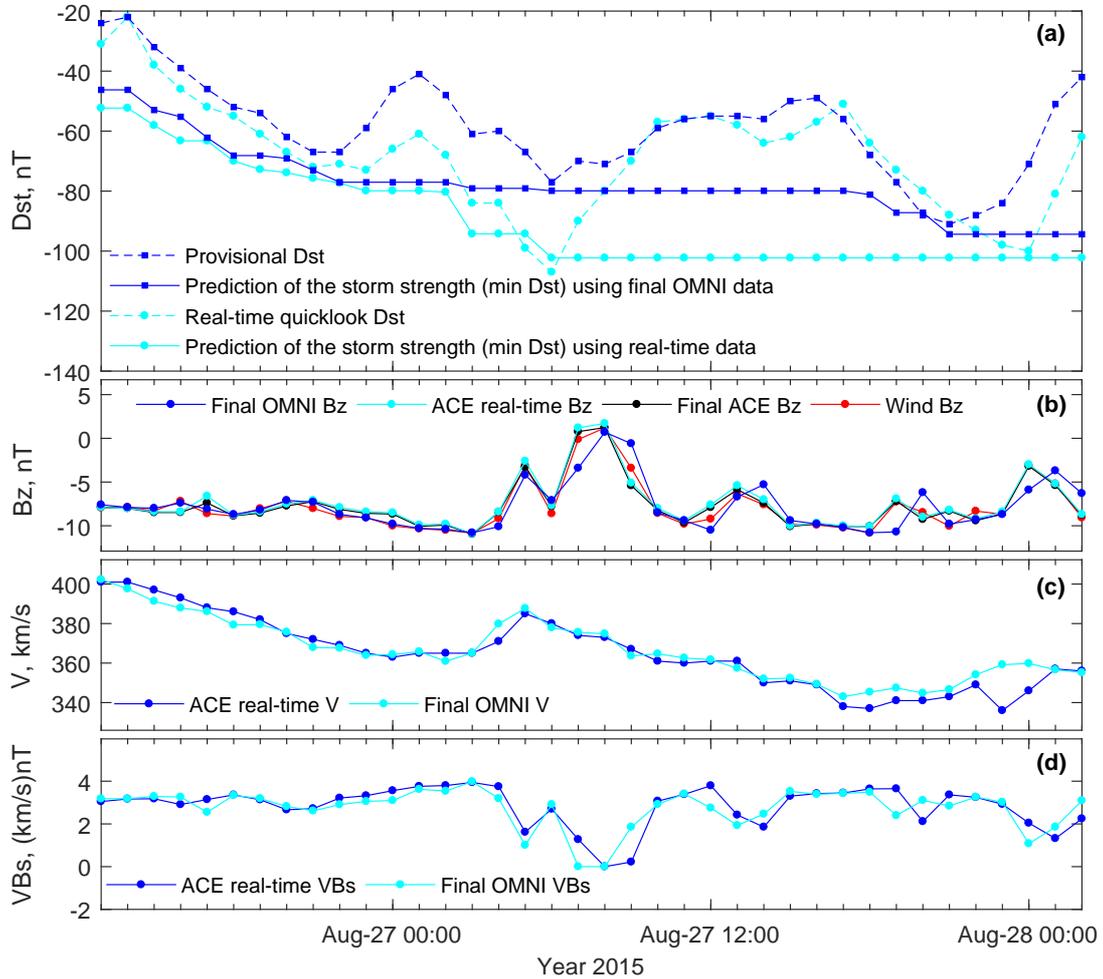}}
\caption{The geomagnetic storm on 26 -- 28 August 2015.
(a) Provisional $Dst$ (dashed blue);
predictions of the peak $Dst_{p}$ using final OMNI input (solid blue);
real-time quicklook $Dst$ (dashed cyan);
predictions of real-time $Dst_{p}$ using real-time input (solid cyan).
(b) Final OMNI IMF $B_{z}$ (blue);
ACE real-time IMF $B_{z}$ (cyan);
Final ACE IMF $B_{z}$ (level 2, black);
Wind $B_{z}$ (not shifted original data, red).
(c) Final OMNI solar wind speed $V$ (cyan);
ACE real-time solar wind speed $V$ (blue).
(d) Final OMNI parameter $VBs$ (cyan);
ACE real-time parameter $VBs$ (blue).}
\label{fig5}
\end{figure}

Figure~\ref{fig5} shows the specific example of
the forecast error due to the difference between quicklook and provisional $Dst$.
Panel (a) gives the provisional $Dst$ and prediction
of its peak $Dst_{p}$ using final input, as well as the same pair for the real-time data.
Panel (b) shows IMF $B_{z}$ from ACE real-time, final OMNI, final ACE, and final Wind.
Final OMNI solar wind speed $V$ and ACE real-time $V$ are in panel (c). Panel (d) has the final OMNI driving parameter $VBs$ and ACE real-time $VBs$.

According to real-time data the storm reached the peak $Dst_{p}$ of -107~nT
at 6:00 UT on 27 August 2015, however the provisional $Dst$ peak
was significantly weaker reaching only -77~nT.
The first ``successful'' real-time forecast ($-81~nT$) of $Dst_{p}$ was issued 4 hours in advance. The maximal prediction of $Dst_{p}$ issued later in real-time within the storm reached --102 nT, producing more accurate prediction of quicklook $Dst_{p}$. However it was 32\% stronger (more negative) than the actual storm strength determined by final $Dst_{p}$.
The primary computational reason for this overestimate was the large difference of about 20~nT between real-time and provisional $Dst$ (the current real-time quicklook $Dst$ index is used in forecast as initial condition).
This storm has three clear intensifications. The two other were at
21:00 UT on 26 August and 21:00 -- 23:00 UT on 27 August, and they were well predicted, since, in particular, the difference between quicklook and provisional $Dst$ was small.
At 7:00 UT on 27 August 2015,
there were also significant differences in $VBs$ between OMNI and individual satellites (Figure~\ref{fig5}b), but the accuracy of prediction was not affected
as it happened one hour after the storm peak.

The deviations between final OMNI and ACE real-time $B_{z}$
were greater than 4 nT in 9\% of cases (Figure~\ref{fig4}b), and the difference between final OMNI and ACE real-time solar wind speed was less than
50 km/s in 97\% of cases (Figure~\ref{fig4}c).
The ACE real-time solar wind density $N$ was mainly underestimated compared
to the final OMNI  (87\% of cases) and the difference was greater
than 5~cm$^{-3}$ nT in 17\% of cases (Figure~\ref{fig4}d).
Thus $B_z$ differences could be much more important than the solar wind
speed differences.
The density errors in real-time are also large. However, in our model
we do not use density as input, since these errors could
overweight the advantages of the $Dst$ models, accounting for Chapman-Ferraro currents.

IMF and solar wind speed differences can come from the several sources:
(1) errors of real-time data, corrected later in the final variant;
(2) gaps in real-time data (filled with the previous values during forecast);
(3) peculiarities of averaging/projection to the bow shock
nose in OMNI and real-time algorithms;
(4) actual differences between measurements of ACE and Wind spacecraft
(over the test period OMNI dataset
is filled with  97\% of Wind data and only 3\% of ACE data).

\begin{figure}
\centerline{\includegraphics[width=1\textwidth,clip=]{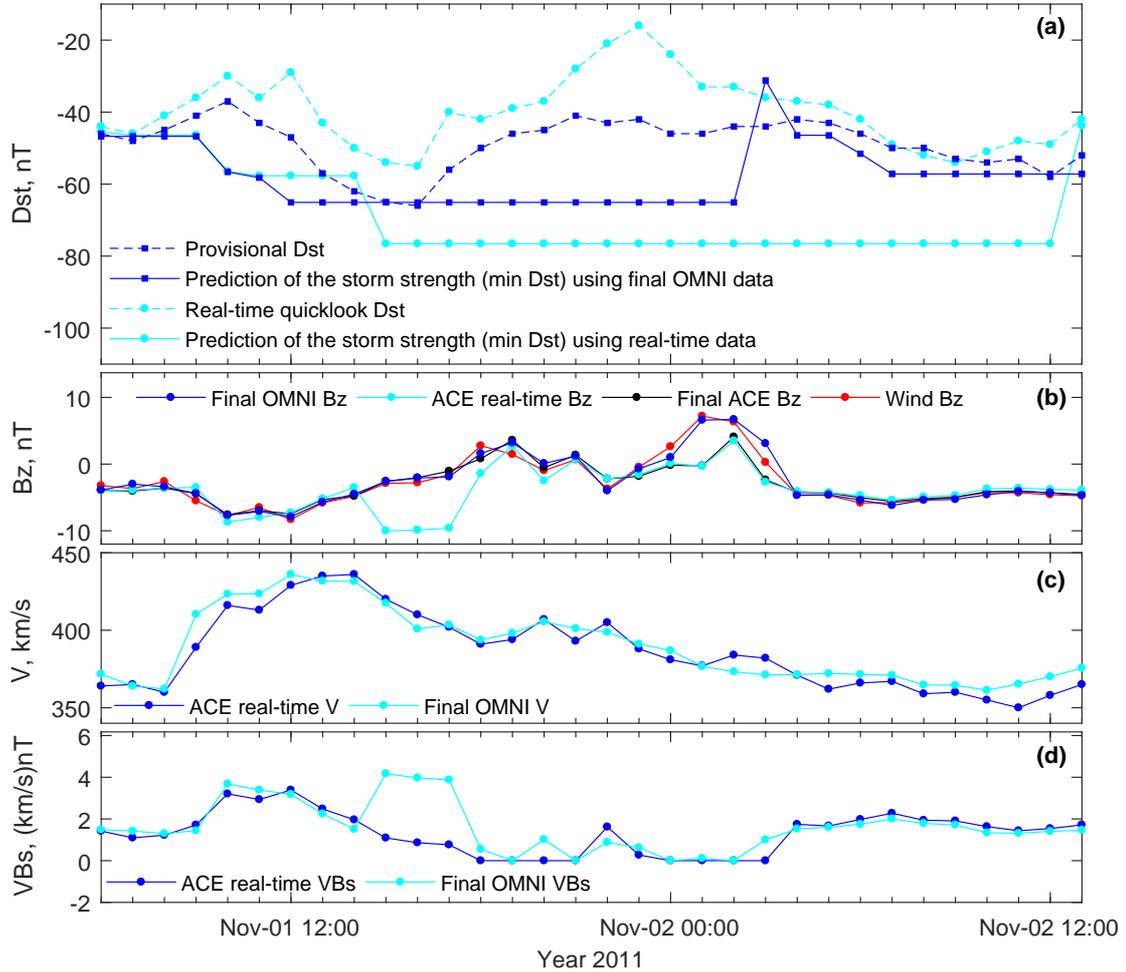}}
\caption{The same as Figure~\ref{fig5} except for
the geomagnetic storm on 1 -- 2 November 2011. }
\label{fig6}
\end{figure}

Figure~\ref{fig6} shows the example of forecast of the geomagnetic storm on 1 -- 2 November 2011 with the error in ACE real-time data.
At 16:00~UT on 1 November 2011 the storm reached the peak $Dst_{p}$
of -66~nT according to the provisional $Dst$. The first ``successful'' forecast was issued 6 hours in advance.
Within this storm the ACE real-time $B_{z}$
was around -10~nT for three hours 15:00--17:00~UT,
while $B_{z}$ from final ACE, Wind, and OMNI were significantly greater (around -2 nT).
Thus the real-time forecast provided the wrong warning at this  intensification.

\begin{figure}
\centerline{\includegraphics[width=1\textwidth,clip=]{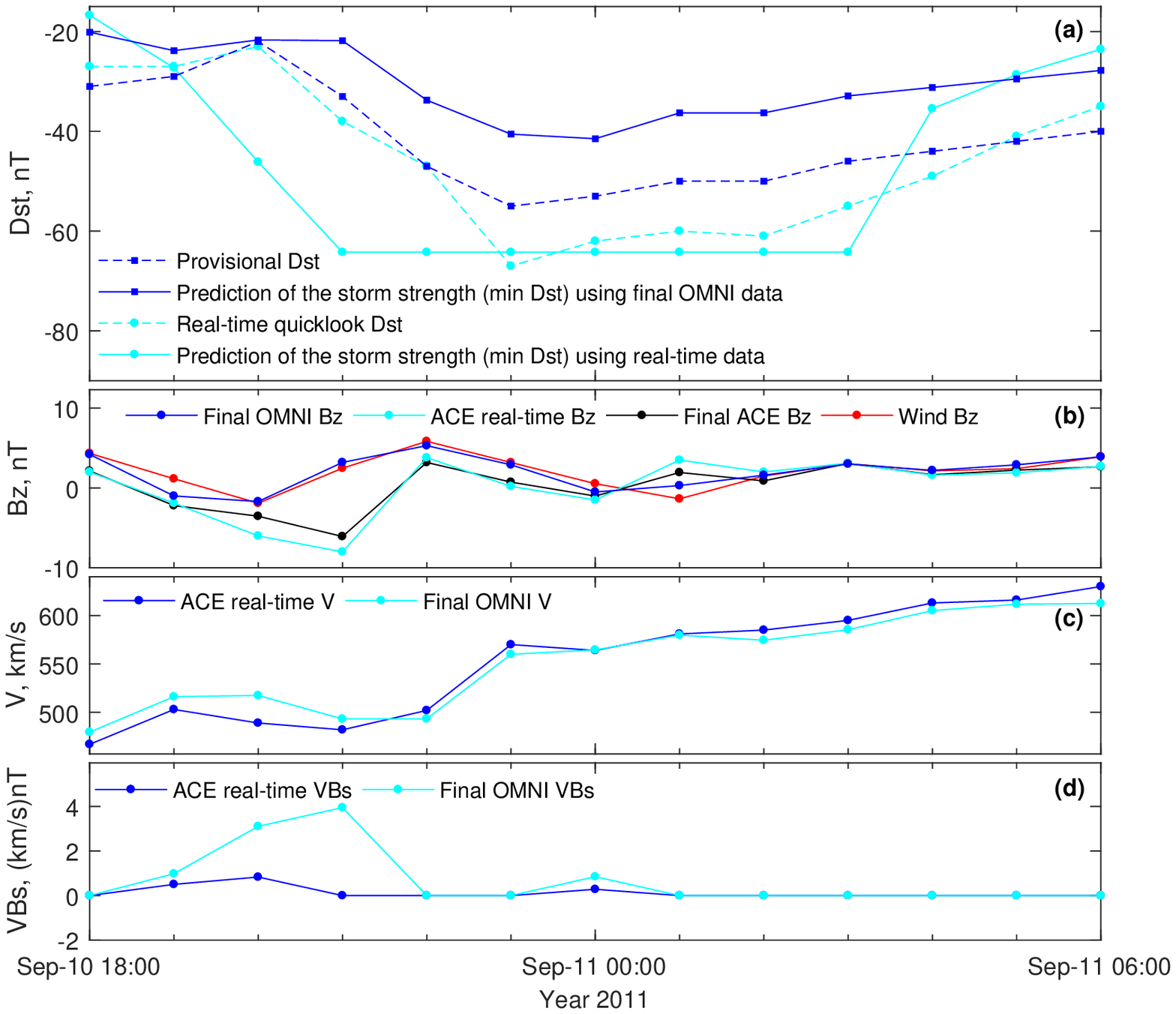}}
\caption{The same as Figure~\ref{fig5} except for
the geomagnetic storm on 11 September 2011.}
\label{fig7}
\end{figure}

\begin{figure}
\centerline{\includegraphics[width=1\textwidth,clip=]{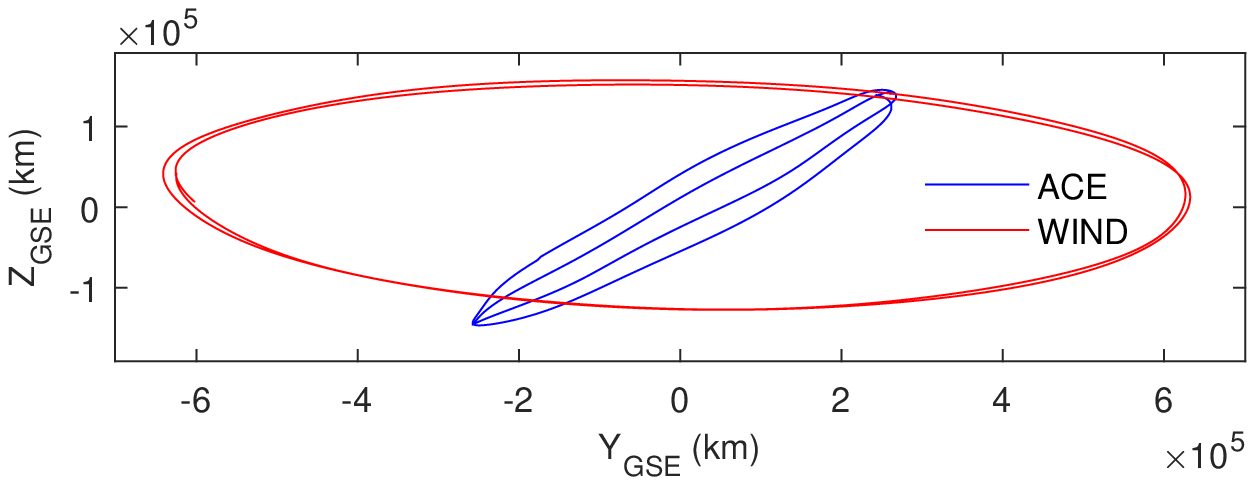}}
\caption{Position of ACE (blue) and Wind (red) in 2011.}
\label{fig8}
\end{figure}

Figure~\ref{fig7} shows the example of ``successful'' forecast on 11 September 2011, using the real-time input, while the forecast with the final OMNI input failed.
At 21:00~UT final OMNI $B_{z}$ and Wind $B_{z}$ (used in OMNI for this storm) were positive around 3~nT, while the ACE real-time and final ACE $B_{z}$ reached the negative value of around --7~nT. The earliest ``successful'' forecast of final peak $Dst_{p}$ was issued 3 hours in advance in case of real-time input data and no warning was issued for the final data variant.
$Dst$ intensification was definitely more consistent with the
ACE observation. Wind spacecraft for years 2011--2016 was in L1 halo orbit with the radius up
to 600 000 km, about twice larger than that for ACE (Figure~\ref{fig8}), thus
using Wind might be less appropriate for the forecast training.

Relatively ubiquitous were single-point (for the hour averaging) discrepancies
between ACE real-time and final OMNI data. They could be caused
by relatively short-duration real-time errors or differences between spacecraft
as well as by specifics of time-shifts and averaging. In terms of forecast quality
such single-point discrepancies were mostly important for the qualification of the ``sudden''
storms with sharp increase of $VBs$.

Table~\ref{table2} summarizes the reasons for the decrease of forecast quality for 49 storms, for which the real-time forecast of $Dst_{p}$ failed, but the forecast of $Dst_{p}$ with the final data was successful. This number includes 21 storms for which
the real-time earliest forecast of $Dst_{p}$ failed, and 28 storms for which maximal predictions of $Dst_{p}$ issued later were overestimated in case of real-time input.
These 49 forecast failures can be definitely ascribed to the errors in the real-time data relative to the final input.
Besides that there were 18 events with the errors
in both the real-time forecast of $Dst_{p}$ and reanalysis with the
final data and 7 events with the
errors only in the runs with the final data. These two latter
types of errors may originate from variety of sources
and are not analyzed here.
\renewcommand{\tabcolsep}{4pt}
\begin{table}
\caption{The summary of reasons for the decrease of forecast quality using real-time input.}
\label{table2}
 \begin{tabular}{@{\extracolsep{5pt}}c cc c c@{}}
  \hline \\[-1.0em]        
          & \multicolumn{2}{c}{\parbox{4.2cm}{\centering Quality decrease for \\the earliest forecast \\in  21 storms}}
					& \parbox{3cm}{\centering Overestimated \\maximal prediction \\in 28 storms}
					& \parbox{4cm}{\centering In total, the decrease \\of forecast quality \\in 49 storms} \\[0.5cm]
					\hline \\ [-2ex]
 & \parbox{2.1cm}{\centering Overestimated \\for 10 storms} & \parbox{2.1cm}{\centering Missed for\\ 11 storms}  & & \\
\\[-2ex] \cline{2-3} \\[-1ex]
$Dst$ overestimated                                             & 6  & -  & 22  & 28 \\[0.5cm]
$VBs$ overestimated                                             & 2  & -  & 2   & 4  \\[0.5cm]
\parbox{3.5cm}{\centering $Dst$ and $VBs$ \\overestimated}      & 2  & -  & 4   & 6  \\[0.5cm]
\parbox{3.5cm}{\centering ACE failure \\$VBs$ underestimated}   & -  & 6  & -   & 6  \\[0.5cm]
$VBs$ underestimated                                            & -  & 4  & -   & 4  \\[0.5cm]
$Dst$ underestimated                                            & -  & 1  & -   & 1  \\[0.1cm]
\hline
\end{tabular}
\end{table}

The main reason of errors (more than half, 28 out of 49) was overestimation of $Dst$ index in real-time. On average quicklook $Dst$ is stronger (more negative) than the provisional index.
In all cases it caused an overestimated forecast. The opposite
situation happened only once, one missed forecast was due to $Dst$ underestimation in real-time.
Four errors with the overestimated forecast were due to $VBs$ overestimation in real-time, while
six such errors were due to both $Dst$ and $VBs$ overestimation.
Four errors with the missed storms were related with $VBs$ underestimate in real-time.
Finally six errors with missed storms were related with ACE data absence in real-time
at the moment of the storm onset. In such a case, the forecast algorithm uses previous
(pre-storm) values of solar wind and IMF, which results in underestimate of $VBs$.

\section{Conclusions}
In this study we review performance of the geomagnetic storm forecasting service StormFocus at \href{http://spaceweather.ru/content/extended-geomagnetic-storm-forecast}{SpaceWeather.Ru} during more than 5 years of operation from July 2011 to December 2016. The service provides the warnings on the expected geomagnetic storm magnitude for the next several hours on an hourly basis. The maximum of the solar cycle 24 was weak, so the most of statistics were rather moderate storms. We verify the quality of forecasts and selection criteria, as well as the reliability of online input data to predict the geomagnetic storm strength in comparison with the final values available in archives.
We also analyze the sources of prediction errors and identified the errors that cannot be removed in real-time and those that can be fixed  by improving the prediction algorithm.

StormFocus service on the basis of ACE real-time and quicklook $Dst$ data issued the successful forecast of peak $Dst_{p}$ (within 25\% of the actual storm magnitude) for 129 out of 153 geomagnetic storms with the detection probability of 0.87.
The algorithm rerun on final OMNI data provided the successful forecast for 148 storms with the detection probability of 0.97.
An important measure of the practical usefulness of the real-time forecast
is the prediction of the actual final $Dst_{p}$ using the real-time input.
It was successful for 130 storms with the detection probability of 0.9. Therefore we confirmed the general reliability of the StormFocus service to provide the advance warnings of geomagnetic storm strength over the period 2011--2016.

Several error sources require special attention. First of all, forecast in real-time operation is somewhat less reliable, than on the final archived data due to errors or imperfections of the real-time input, both solar wind, IMF and $Dst$.
Such reasons accounted for more than a half of all forecast errors.
In particular such reasons most likely explain a strong increase of a number of later overestimates (after successful forecast was given) from $\sim$10\% to 24\%, when the real-time data are compared with the final index.

Real-time errors are often overlooked during performance analysis, since
they are nominally beyond control during an algorithm design. One can advise to verify the
forecast quality additionally on the real-time data streams if available. Statistics on expected
differences between real-time and final data would be also very helpful.
Alternatively one may use the forecast models with the probabilistic output
(providing the range of possible values with some degree of certainty), however
such models require much larger data amounts for training.

The second substantial error was related with the generation of the ``sudden'' storm warning, based on a sharp increase of $VBs$. In real-time about a half of the ``sudden'' storms were missed and several false warnings were
generated. The primary reason was again related with the differences between real-time
and final data. To remediate such errors we additionally introduced in the algorithm
a condition, canceling the warning in a case it was caused by a single-point peak in $VBs$.
However, it should be noted, that the sudden storm variant of the forecast was designed
to warn on very strong storms (as it was during period 1995--2010, used for the design).
Stronger storms are usually related with more stable solar wind and IMF input and develop
faster, thus extrapolations in our forecast are more reliable.
During the test period the storms were much weaker on average, practically all ``sudden''
storms were weaker (less negative) than -150 nT. For such events ``sudden'' storm warnings are expected to be less reliable.

In conclusion, the geomagnetic storm forecasting service StormFocus that provides the warnings of future geomagnetic storm magnitude proved to be quite successful after more than five years of online operation and can be recommended for applications. A few of the most registered forecast failures were caused by the errors in the real-time input data (relative to final data, appearing later). This source of errors needs special attention during the forecast algorithm design.

\begin{appendix}
\section{The prediction of storm strength}
The prediction technique of the peak $Dst_{p}$ during geomagnetic storms is based
on the differential equation of the $Dst$ index evolution introduced by \citet{Burton1975}
as a functional relation between solar wind and $Dst$.
\begin{equation} \label{Eq_Burton}
\frac{dDst^*}{dt}=Q(t)-\frac{Dst^*}{\tau(t)}
\end{equation}
Here $Q$ (mV/m) is the solar wind input function. We use the model variant by \citet{OBrien2000b},
where the parameter $Q$ is linearly connected with the driving parameter $VBs$
\begin{equation}
Q(t)=-4.4(VBs-0.5) \\
\tau(t)=2.4e^\frac{9.74}{4.69+VBs}
\end{equation}
\begin{equation}
 VBs = \left\{ \begin{array}{ll}
         |VB_{z}|, & \mbox{$B_z < 0$}, \\
         0, & \mbox{$B_z \geq 0$}.\end{array} \right.
\end{equation}
\begin{equation} \label{Eq_Dst_cor}
Dst^*=Dst-7.26\sqrt{P_{dyn}}+11.
\end{equation}
Here $V$ (km/s) is solar wind plasma speed, $B_s$  (nT) is
southward component of IMF in GSM, $P_{dyn}$ (nPa) is solar wind dynamic pressure,
and $\tau$ (hours) is  decay time constant, associated with the loss
processes in the inner magnetosphere.
$Dst^*$ is the pressure corrected $Dst$ index,
from which the effects of the magnetopause currents have been removed.
However, we neglect the difference between $Dst$ and $Dst^*$, and do not compute
this pressure correction, because generally it is rather small,
and often real-time solar wind density  is  quite different from the
calibrated data, appearing later in the archives.

To predict future strength of a storm we use the  solution of the differential
equation~(\ref{Eq_Burton}), with an initial condition (solar wind, IMF and
quick-look $Dst$) taken at some ``zero'' moment
and assuming stationary solar wind input, equal to that at the
initial point (hereafter $Q_0$, $\tau_0$).
\begin{equation} \label{Eq_Sol}
Dst(t)=e^{-\frac{t}{\tau_0}}\cdot(Dst(0)-Q_0\cdot\tau_0)+Q_0\cdot\tau_0
\end{equation}

This solution is the monotonously decreasing function of time,
and when $t\rightarrow\infty$, it approaches the steady-state value.

We compute two variants of the forecast $Dst$.
The storm saturation level reached at the steady-state solution
is used as a prediction of the lower limit of peak $Dst_{p}$.

\begin{equation} \label{Eq_Sat}
Dst(\infty)=\lim\limits_{t\to \infty }Dst(t)=Q_0\cdot\tau_0
\end{equation}

To predict the upper limit we select the intermediate point
on the saturation trajectory determined by the solution of the
discrete variant of equation~(\ref{Eq_Burton}) given by

\begin{equation} \label{Eq_Disc}
Dst(k+1)=Dst(k)-\frac{1}{\tau(k)}Dst(k)+Q(k)
\end{equation}
The solution of this equation $k$ hours ahead for constant $Q$ and $\tau$ is
\begin{equation} \label{Eq_Sol_Disc}
{Dst}(k+1)=\left(1-\frac{1}{\tau_0}\right)^{k+1}Dst(0)+Q_0\sum_{i=0}^{k}\left(1-\frac{1}{\tau_0}\right)^{i}
\end{equation}
As the upper limit we use the three-hours-ahead extrapolation, which was justified by the statistics.
\begin{equation} \label{Eq_Sol_3hour}
\widehat{Dst}(+3)=\left(1-\frac{1}{\tau_0}\right)^3Dst(0)+\sum_{i=0}^{2}\left(1-\frac{1}{\tau_0}\right)^{i}Q_0
\end{equation}

$\widehat{Dst}(+3)$ is computed routinely every hour, and to avoid its excessive variability
the final prediction $\overline{Dst}$ is kept at the minimum of all $\widehat{Dst}(+3)$
obtained after the last ``storm end'' flag. The ``storm end'' is signaled and $\overline{Dst}$ returns to the current $\widehat{Dst}(+3)$, when IMF $B_z\ge 1$ nT during 3 hours or IMF $B_z\ge -1.8$ nT during 11 hours.

The prediction of both upper and lower limits, $\overline{Dst}$ and  $Dst(\infty)$, of the peak $Dst_{p}$ is proved to be useful only for a group of larger ''sudden`` storms, associated with a sufficiently sharp increase of $VBs$. All other storms are then called “gradual”.

We introduced the following criteria for the sharp increase of $VBs$: $I(k)=VBs(k)-VBs(k-1)$:
\begin{equation} \label{Eq_Jump1}
I(k)>4.4 \,\, and \, \sum_{j=k-2}^{k}I_{j}>5.7
\end{equation}
\begin{equation} \label{Eq_Jump2}
VBs(k)>6.2 \, or \, (VBs(k)>5.5 \, and \, I(k-1)>0)
\end{equation}

The thresholds in (\ref{Eq_Jump1}, \ref{Eq_Jump2}) are given in mV/m. Condition~(\ref{Eq_Jump1}) checks for
the sharp increase of $VBs$ during last three hours and the last hour in particular. Condition~(\ref{Eq_Jump2})
requires that $VBs$ is sufficient to cause a strong enough storm. Numerical coefficients in~(\ref{Eq_Jump1}, \ref{Eq_Jump2})
were selected empirically  on historical
data to perform the best possible forecast.

We issue a forecast of future ``sudden'' storm strength at a specific moment of $VBs$ jump
when both criteria (\ref{Eq_Jump1}, \ref{Eq_Jump2}) are fulfilled. We expect a stable prolonged
storm-grade solar wind input and define both a lower limit $Dst(\infty)$ and an upper limit $\widehat{Dst}(+3)$,
so that the real peak value is expected to be between $Dst(\infty)<Dst_{p}<\overline{Dst}$.
For weaker storms with $VBs < 10.9$ mV/m, when $Dst(\infty)<-150$~nT, we use $Dst(\infty)=0.85 Q_0\cdot\tau_0$, since such storms require too long time to approach their saturation value. This two-level forecast works best for the
strongest storms, with shorter saturation times (proportional to $\tau$ in Burton equation)

Gradually developing storms (for which ``sudden'' criterion is not fulfilled) usually has longer
saturation times, while variability of the input
is larger, thus such events are rarely reaching expected saturation level.
For such events the prediction of peak $Dst_{p}$ is performed on the basis
of three-hour forecast $\overline{Dst}$ only. The storm alert is issued when
the forecast of peak $Dst_{p}$ reaches the threshold of -50~nT.

Reviewing the performance of the algorithm during more than 5 years of operation, we partially modified
the criteria to detect ``sudden'' storms, associated with a sharp increase of $VBs$.
Single-point outliers in real-time $VBs$ are actually observed more often than in
final data. To minimize the risk of false ``sudden'' storm warnings, the jump criteria
(\ref{Eq_Jump1},\ref{Eq_Jump2}) are now augmented with the rule, cancelling
the ``sudden'' storm warning if it was generated
due to a single-point outlier in $Q$:
 The lower limit is removed from the forecast, when $VBs \leq 3$ at step $k$,
$VBs<5.5$ over two hours at steps $k-2$ and $k-3$,
and the conditions (\ref{Eq_Jump1}, \ref{Eq_Jump2}) were fulfilled at previous step $k-1$.
Over the period of July 2011 -- December 2016 there were 5 such cases.

\end{appendix}

\begin{acknowledgements}
The authors are grateful to the National
Space Science Data Center for the OMNI 2 database, to the National
Oceanic and Atmospheric Administration (NOAA) for ACE real-time
solar wind (RTSW) data, to the WDC for Geomagnetism (Kyoto) for
the $Dst$ index data. We thank the referee for valuable comments on this study.
The work was supported by Russian Science Foundation, N16-12-10062.
\end{acknowledgements}




\end{document}